\documentclass[12pt]{article}
\usepackage[plainpages=false,pdfpagelabels,unicode]{hyperref}
\usepackage[english]{babel}
\usepackage{comment}
\usepackage{amsfonts}
\usepackage{graphicx}
\usepackage{epstopdf}
\usepackage{epsfig}
\usepackage{amssymb}
\usepackage{setspace}
\usepackage{caption}
\usepackage{mathrsfs}
\usepackage{mathtools}
\usepackage{color}
\usepackage{amsmath}
\usepackage{amsthm}
\usepackage{float}
\usepackage[utf8]{inputenc}
\usepackage[square,comma,numbers,sort&compress]{natbib}

\newcommand {\be}{\begin{equation}}
\newcommand {\ee}{\end{equation}}

\newcommand {\bea}{\begin{array}}

\newcommand {\eea}{\end{array}}

\newcommand {\dt}{\text{d}t}
\newcommand {\dr}{\text{d}r}
\newcommand {\dx}{\text{d}x}
\newcommand {\dphi}{\text{d}\phi}
\newcommand{\dd}{\text{d}}

\numberwithin{equation}{section}

\begin{document}
\begin{titlepage}
\vspace{1cm} 
\begin{center}
{\Large \bf {Separability of the Hamilton-Jacobi equation and equatorial circular photon orbit in Kerr-Sen-Taub-NUT spacetime}}\\
\end{center}
\vspace{1cm}
\begin{center}
\renewcommand{\thefootnote}{\fnsymbol{footnote}}
Delvydo Melvernaldo{\footnote{email: delvydo@zoho.com or 7216007@student.unpar.ac.id}}\\
Center for Theoretical Physics, Department of Physics,\\
Parahyangan Catholic University, Bandung 40141, Indonesia\\
\renewcommand{\thefootnote}{\arabic{footnote}}
\end{center}

\begin{abstract}
In this paper, we show the additive separation of the Hamilton-Jacobi equation, present the 4-velocity of the test particles, and attempt to find the equatorial circular photon orbit (ECPO) in the Kerr-Sen-Taub-NUT (KSTN) solution of the low energy limit of heterotic string theory. In the process of trying to find the ECPO, we instead encounter a ring singularity that could be outside the horizon of the KSTN black hole.
\end{abstract}
\end{titlepage}
\onecolumn
\bigskip 

\section{Introduction}
\label{sec:intro}
The Kerr-Sen-Taub-NUT (KSTN) spacetime describes a rotating electrically charged mass with NUT parameter as predicted in the low energy limit of heterotic string theory which is analogous to the Kerr-Newman-Taub-NUT solution of Einstein-Maxwell theory. The author of \cite{Siahaan_2020_KSTN} employed the Hassan-Sen \cite{Hassan_1992} transformation on the Kerr-Taub-NUT (KTN) solution to get the KSTN solution. The Hassan-Sen transformation has been used to map any stationary and axial symmetric spacetime that solves the vacuum Einstein equation to a new solution in the low energy limit of heterotic string theory. The Kerr-Sen solution, which is the solution one would get if the NUT parameter of the KSTN spacetime vanishes, was obtained by Ashoke Sen \cite{Sen_1992} by applying the Hassan-Sen transformation on the Kerr solution.

In the process of analyzing the behavior of a test particle in a spacetime, the Hamilton-Jacobi formulation is often used to obtain the equations of motion of the test particle \cite{dewitt1973black}. This formulation is particularly useful in finding a conserved quantity in the equations of motion which then can be utilized to simplify the equations. The additive separability of the Hamilton-Jacobi equation (HJE) is critical in finding these conserved quantities. However, Ref. \cite{Siahaan_2020_KSTN} noted that the HJE isn't likely to be separable due to the presence of a cross-term in the denominator. This impedes the process of calculating the behavior of a test particle in this spacetime.

The organization of this paper is as the followings. In Sect. \ref{sec:KSTN}, we provide a short summary of the KSTN solution. In Sect. \ref{sec:HJE-EOM}, we separate the HJE of the KSTN solution, find the corresponding conserved quantity, and obtain the 4-velocity of electrically neutral test particles. In Sect. \ref{sec:ECPO}, we attempt to find the equatorial circular photon orbit.

\section{Kerr-Sen-Taub-NUT spacetime}
\label{sec:KSTN}
Using coordinates $x^\mu = [t,r,x,\phi]$ where $x = \cos\theta$ in Boyer-Lindquist coordinates, the metric of KSTN spacetime in Einstein frame can be expressed as \cite{Siahaan_2020_KSTN}
\begin{equation}
\label{eq:KSTN-ds2}
\begin{split}
ds^2 = & -\frac{\Xi}{\Sigma} \left[ \dt - \frac{2M (\Delta_r l (1-x) - a \Delta_x ((M-b)r + l^2 + al) ) }{(M-b)\Xi} \dphi \right]^2\\
& + \Sigma \left[ \frac{\dr^2}{\Delta_r} + \frac{\dx^2}{\Delta_x} + \frac{\Delta_r \Delta_x \dphi^2}{\Xi} \right]
\end{split}
\end{equation}
with
\begin{equation}
    \begin{split}
		\Sigma &= r(r+2b) + (l+ax)^2 + \frac{2bl(l+ax)}{M-b},\\
		\Xi &= r^2 - 2(M-b)r + a^2 x^2 - l^2,\\
		\Delta_r &= r^2 - 2(M-b)r + a^2 - l^2,\\
		\Delta_x &= 1-x^2,
    \end{split}
    \label{eq:sigma-delta-chi}
\end{equation}
where $M$ is the mass of the gravitating object, $a$ is the rotational parameter, $l$ is the NUT parameter, and the electric charge $Q$ is represented by $b$ with the relation $b = Q^2 / (2M)$.

The low energy effective action for heterotic string in Einstein frame used for this solution is \cite{Johnson_1994}
\begin{equation}
S = \int d^4x \sqrt{-g} \left[ R(g) - \frac{1}{2} \nabla_\mu \Phi \nabla^\mu \Phi - \frac{e^{-\Phi}}{8} F_{\mu\nu} F^{\mu\nu} - \frac{e^{-2\Phi}}{12} H_{\mu\nu\lambda} H^{\mu\nu\lambda} \right]
\end{equation}
with \cite{Siahaan_2020_KSTN}
\begin{equation}
\label{eq:fields}
\begin{split}
A_\mu \dx^\mu = &\ \frac{\sqrt{2}Q}{\sqrt{\Sigma}(M-b)} \left[ (alx+l^2+(M-b)r ) \dt - a \Delta_x ((M-b)r+l^2+al) \dphi \right], \\
H_{\alpha\beta\mu} = &\ \partial_\alpha B_{\beta\mu} + \partial_\mu B_{\alpha\beta} + \partial_\beta B_{\mu\alpha} - \frac{1}{4} (A_\alpha F_{\beta\mu} + A_\mu F{\alpha\beta} + A_\beta F_{\mu\alpha} ),  \\
\Phi = & -2 \ln \sqrt{\frac{\Sigma}{r^2+(l+ax)^2)}},
\end{split}
\end{equation}
where $A_\mu$ is the Maxwell field which gives the field strength tensor $F_{\mu\nu} = \partial_\mu A_\nu - \partial_\nu A_\mu2$, $\Phi$ is the dilaton field, $B_{\mu\nu}$ is the second-rank anti-symmetric tensor field, and $H_{\alpha\beta\mu}$ is a Chern-Simons term in the absence of $B_{\mu\nu}$. The non-zero components of $B_{\mu\nu}$ is
\begin{equation}
B_{t\phi} = -B_{\phi t} = \frac{2b}{\sqrt{\Sigma}(M-b)}\left[ l \Delta_r (1-x) - a \Delta_x [(M-b)r + l^2 + al] \right].
\end{equation}
Setting $l=0$ reduces the set of fields solutions (\ref{eq:fields}) to that of Kerr-Sen \cite{Sen_1992} and the solution becomes the Kerr-Taub-NUT solution after setting $Q=0$.

Just as Taub-NUT or Kerr-Taub-NUT solutions have conical singularities at the poles where the metric tensor becomes non-invertible \cite{Ong_2017}, the metric (\ref{eq:KSTN-ds2}) also has such singularity at $x=-1$. Whether or not the KSTN spacetime has well defined black holes due to this conical defect, one can still find the metric singularities at $\Delta_r = 0$, i.e.
\begin{equation}
r_\pm = M-b \pm \sqrt{(M-b)^2 + l^2 - a^2}.
\end{equation}
The frame dragging effect can also be seen in this spacetime where the angular velocity of an observer with constant $r$ and $x$ is
\begin{equation}
\label{eq:Omega}
\Omega = - \frac{g_{t\phi}}{g_{\phi\phi}} = \frac{-2M\Xi(M-b) \left[ a\Delta_x [(M-b)r+l^2+al]-l\Delta_r(1-x) \right]}{4M^2 \left[ a\Delta_x [(M-b)r+l^2+al]-l\Delta_r(1-x) \right]^2 - \Sigma^2 (M-b)^2 \Delta_r \Delta_x}
\end{equation}
which at the outermost metric singularity $r_+$ reduces to
\begin{equation}
\Omega_+ = \frac{a(M-b)}{2M(r_+(M-b)+l^2+al)}.
\end{equation}
The area covered by the sphere with radius $r_+$ can be computed as
\begin{equation}
\mathcal{A} = \int^{2\pi}_0 \int^1_{-1} \sqrt{g_{xx} g_{\phi\phi}}\ \dx\ \dphi = \frac{8 \pi M \left[r_+(M-b) + l^2 + al \right]}{M-b}.
\end{equation}

\section{Hamilton-Jacobi equation separation and equations of motion}
\label{sec:HJE-EOM}
Let us consider the Lagrangian
\begin{equation}
2 \mathscr{L} = g_{\mu\nu} \dot{x}^\mu \dot{x}^\nu
\end{equation}
where the overdot denotes differentiation with respect to an affine parameter $\lambda$ with the normalization condition is $g_{\mu\nu} \dot{x}^\mu \dot{x}^\nu = \delta$ where one can substitute $\delta=0$ for the null case and $\delta = -1$ for the timelike case. The Hamilton-Jacobi equation (HJE) is
\begin{equation}
\label{eq:generalHJE}
g^{\mu\nu} \partial_\mu S \partial_\nu S = \delta
\end{equation}
with the Jacobi action ansatz
\begin{equation}
\label{eq:S-ansatz}
S = -Et + L\phi + S_r(r) + S_x(x),
\end{equation}
where $E$ and $L$ are respectively the energy and angular momentum for massless particles but are respectively the specific energy and specific angular momentum for timelike particles. Here, $S_r$ and $S_x$ are respectively arbitrary functions that depend only on $r$ and $x$. Also notice that
\begin{equation}
\label{eq:partial S=xdot}
\partial_\mu S = P_\mu = \dot{x}_\mu.
\end{equation}
With that relation, we have
\begin{equation}
\label{eq:E}
E = -g_{t\mu} \dot{x}^\mu = \frac{\Xi}{\Sigma} \dot{t} - \frac{2M \left[ \Delta_r l (1-x) -a\Delta_x((M-b)r+l^2+al) \right] }{\Sigma (M-b)} \dot{\phi},
\end{equation}
and
\begin{equation}
\label{eq:L}
\begin{split}
L = g_{\phi\mu} \dot{x}^\mu = & \frac{2M \left[ \Delta_r l (1-x) -a\Delta_x((M-b)r+l^2+al) \right] }{\Sigma (M-b)} \dot{t} \\
& + \left[\frac{\Sigma\Delta_x\Delta_r}{\Xi} - \frac{4M^2 \left[ \Delta_r l (1-x) -a\Delta_x((M-b)r+l^2+al) \right]^2}{\Sigma \Xi (M-b)^2} \right] \dot{\phi}.
\end{split}
\end{equation}
Solving equations (\ref{eq:E}) and (\ref{eq:L}), we obtain
\begin{equation}
\label{eq:tdot}
\begin{split}
\dot{t} =& \frac{E}{\Xi}\left[ \Sigma - \frac{4M^2 \left[ \Delta_r l (1-x) -a\Delta_x((M-b)r+l^2+al) \right]^2 }{\Sigma\Delta_r\Delta_x(M-b)^2} \right] \\
&+ \frac{2M \left[ \Delta_r l (1-x) -a\Delta_x((M-b)r+l^2+al) \right]}{\Sigma\Delta_x\Delta_r(M-b)}L,
\end{split}
\end{equation}
and
\begin{equation}
\label{eq:phidot}
\dot{\phi} = \frac{\Xi L}{\Sigma \Delta_x \Delta_r} - \frac{2M \left[ \Delta_r l (1-x) -a\Delta_x((M-b)r+l^2+al) \right] }{\Sigma\Delta_x\Delta_r(M-b)} E.
\end{equation}

With the Jacobi action ansatz (\ref{eq:S-ansatz}), the metric (\ref{eq:KSTN-ds2}), and the additive separation of variables algorithm explained in appendix \ref{app}, the HJE (\ref{eq:generalHJE}) becomes
\begin{equation}
\begin{split}
\Sigma\ \delta = & \Delta_r \left(\partial_r S_r(r) \right) ^2 + \Delta_x \left( \partial_x S_x(x) \right)^2 + \left[ \frac{1}{\Delta_x} - \frac{a^2}{\Delta_r} \right] L^2 \\
& - 2 \left[ \frac{F_x}{\Delta_x} - \frac{F_r}{\Delta_r} \right] EL + \left[ \frac{\psi_x}{\Delta_x} - \frac{\psi_r}{\Delta_r} \right] E^2
\end{split}
\end{equation}
where\footnote{The second term in both $F_x$ and $F_r$ cancels out in the HJE. This is just for convenience so that $F_x=0$ at the equator $x=0$.}
\begin{equation}
\begin{split}
F_x = & \frac{2Ml(1-x)}{M-b} - \Delta_x \left[ \frac{2Ml}{M-b} \right],\\
F_r = & \frac{2Ma [(M-b)r+l^2+al]}{M-b} - \Delta_r \left[ \frac{2Ml}{M-b} \right],\\
\psi_x = & \frac{8M^2l^2(1-x)x}{(M-b)^2} - \Delta_x \left[ a^2x^2 + \frac{4Malx}{M-b} \right] ,\\
\psi_r = & 4M^2 \left[ 2r \left( M-b + \frac{l^2+al}{M-b} \right) - (a^2 - l^2) + \left( \frac{a+l}{M-b} \right)^2 l^2 \right] \\
&+ \Delta_r \left[ r^2 + 2r(M+b) + 4M^2 - \left(\frac{M+b}{M-b}\right)^2 l^2 \right].
\end{split}
\end{equation}
The separability of the HJE implies the existence of a conserved quantity termed as the Carter constant $K$ \cite{Carter_1968b} where\footnote{These particular expressions are chosen such that $K=0$ if $x=\dot{x}=0$.}
\begin{equation}
\Delta_x (\partial_x S_x)^2 + \frac{x^2}{\Delta_x}L^2 - \frac{2F_x}{\Delta_x} EL + \frac{\psi_x}{\Delta_x}E^2 - \left[ \frac{2Malx}{M-b} + a^2 x^2 \right] \delta = K,
\end{equation}
and
\begin{equation}
\Delta_r (\partial_r S_r)^2 + \left[1 - \frac{a^2}{\Delta_r}\right] L^2 + \frac{2F_r}{\Delta_r} EL - \frac{\psi_r}{\Delta_r}E^2 - \left[ r(r+2b) + \frac{l^2 (M+b)}{M-b} \right] \delta = -K.
\end{equation}
Using the relation (\ref{eq:partial S=xdot}), we get
\begin{align}
\label{eq:xdot}
\Sigma \dot{x} = \pm \sqrt{\mathcal{X}},\\
\label{eq:rdot}
\Sigma \dot{r} = \pm \sqrt{\mathcal{R}},
\end{align}
with
\begin{align}
\label{eq:fancyX}
\mathcal{X} = \mathcal{X}(x) = & - x^2 L^2 + F_x EL - \psi_x E^2 + \Delta_x \left[ K + \left( \frac{2Malx}{M-b} + a^2 x^2 \right) \delta \right],\\
\label{eq:fancyR}
\mathcal{R} = \mathcal{R}(r) = & a^2 L^2 - F_r EL + \psi_r E^2 - \Delta_r \left[ K + L^2 - \left( r(r+2b) + \frac{l^2(M+b)}{M-b} \right) \delta \right].
\end{align}
The $\pm$ signs in equations (\ref{eq:xdot}) and (\ref{eq:rdot}) are independent from each other but must stay consistent once the signs are chosen. The $+$($-$) sign in equation (\ref{eq:rdot}) denotes the outgoing(ingoing) geodesics. 
Just as in \cite{Cebeci_2019,Zakharov_1994,Mino_2003}, one can introduce a new time parameter $\sigma$ such that
\begin{equation}
\frac{\dd \sigma}{\dd \lambda} = \frac{1}{\Sigma}.
\end{equation}
With this new parameter and equations (\ref{eq:xdot}) and (\ref{eq:rdot}), we can use $\mathcal{R}$ and $\mathcal{X}$ as the effective potentials in $r$ and $x$
\begin{align}
\label{eq:EffPot_r}
\left( \frac{\dd r}{\dd \sigma} \right)^2 = &\ \mathcal{R}(r)  ,\\
\label{eq:EffPot_x}
\left( \frac{\dd x}{\dd \sigma} \right)^2 = &\ \mathcal{X}(x) .
\end{align}
The effective potential $\mathcal{R}(r)$ can also be written in the polynomial form
\begin{equation}
\mathcal{R}(r) = f_4 r^4 + f_3 r^3 + f_2 r^2 + f_1 r + f_0
\end{equation}
where
\begin{align}
f_4 = &\ E^2+\delta,\\
f_3 = &\ 2 \left[ 2bE^2 - (M-2b) \delta \right],\\
\begin{split}
f_2 = &\ E^2 \left(4b^2-2l^2+a^2\right) + 4lLE - K - \frac{4bl^2E^2}{M-b} - \left[ \frac{2blE}{M-b}-L \right]^2\\
&\ + \left[ a^2 + \frac{2bl^2}{M-b} - 4b(M-b) \right] \delta,
\end{split}\\
\begin{split}
f_1 = &\ 2(M-b)\left[ (L-aE-2El)^2 + K \right] + 4b \left[ a^2E^2 + (4al+5l^2)E^2 - (a+2l)LE \right] \\
&\ + \frac{8b^2l(a+2l)E^2}{M-b} + 2 \left[ b(a^2-3l^2) - l^2(M-b) \right] \delta,
\end{split}\\
\begin{split}
f_0 = &\ \left[ (a+l)(3a+5l)E^2 - 4(a+l)LE +L^2+K \right]l^2  - Ka^2 + \frac{8b^2l^3(a+l)E^2}{(M-b)^2}\\
&\  + \frac{4bl^2E(a+l)}{M-b} \left[ 3lE + aE - L \right] + l^2 (a^2 - l^2) \left[ \frac{M+b}{M-b} \right] \delta.
\end{split}
\end{align}
By looking at the coefficient of the highest order of $r$, notice that the motion can be unbounded ($r \to \infty$) if $E^2 > -\delta$. For $E^2 < -\delta$, the motion is bounded and cannot diverge to infinity. For the case $E^2 = -\delta $, whether the motion is finite or infinite depends on the other parameters for the allowed regions and is said to be marginally bounded.

\section{Equatorial circular photon orbit}
\label{sec:ECPO}
Since we will be discussing photons, let us immediately use $\delta=0$ for the rest of this section. I will first consider the constraints necessary for a photon to move over the equatorial plane $x=0$, namely\footnote{The notation $X|_y$ means the function $X$ is evaluated at some fixed $y$.}
\begin{equation}
\label{eq:X_x=0}
\mathcal{X}|_{x=0}=K=0
\end{equation}
and
\begin{equation}
\label{eq:dX_x=0}
\left. \frac{\dd \mathcal{X}}{\dd x} \right|_{x=0} = \frac{\left[ 8M^2 l E^2 + 4ME(aE-L) \right] l}{(M-b)^2} = 0.
\end{equation}
One can see that the constraint (\ref{eq:dX_x=0}), which corresponds to the acceleration in $x$, is satisfied for zero NUT parameter. For non-zero NUT parameter, we have an additional constrain for the energy and angular momentum of the particle.
The value of the particle's angular momentum $L$ that satisfies the latter constraint, i.e. Eq. (\ref{eq:dX_x=0}), is
\begin{equation}
\label{eq:L-eq}
L = aE + \frac{2EMl}{M-b}.
\end{equation}
I will use these values for $K$ and $L$ from equations (\ref{eq:X_x=0}) and (\ref{eq:L-eq}) for the rest of this section. Since the inequality
\begin{equation}
\left. \frac{\dd^2 \mathcal{X}}{\dd x^2} \right|_{x=0} = -\frac{8M^2 l^2 E^2}{(M-b)^2} < 0
\end{equation}
is satisfied, this motion over the equatorial plane is stable, i.e. small perturbations in the $x$ direction will not throw the particle away from the equator.

Next, let us consider the constraints for circular motion, namely
\begin{equation}
\label{eq:R_c=0}
\mathcal{R}|_{r=r_c} = E^2 \left[ r_c(r_c+2b) + \frac{l^2(M+b)}{(M-b)} \right]^2 = 0
\end{equation}
and
\begin{equation}
\label{eq:dR_c=0}
\left. \frac{\dd \mathcal{R}}{\dd r} \right| _{r=r_c} = 4E^2 (r_c+b) \left[ r_c(r_c+2b) + \frac{l^2(M+b)}{M-b} \right] = 0.
\end{equation}
Notice that the expression inside the bracket on both equations (\ref{eq:R_c=0}) and (\ref{eq:dR_c=0}) is in fact $\Sigma|_{x=0,r=r_c}$. From the Kretchmann scalar $R_{\alpha\beta\mu\nu} R^{\alpha\beta\mu\nu}$ at the equatorial plane, which is
\begin{equation}
\left. R_{\alpha\beta\mu\nu} R^{\alpha\beta\mu\nu} \right|_{x=0} = \frac{\mathcal{O}(r^6)}{ \left[ (M-b) \Sigma|_{x=0} \right]^6},
\end{equation}
one can see that $\Sigma|_{x=0} = 0$ would make the curvature scalar singular.
Looking into this further, if we solve Eq. (\ref{eq:R_c=0}) for $r_c$ with $E\neq0$ regardless of the curvature singularity, we get
\begin{equation}
\label{eq:r_c}
r_c = -b \pm \sqrt{b^2 - \frac{l^2(M+b)}{M-b}}.
\end{equation}
It is interesting to notice that $r_c$ is positive if and only if $M<b$ and this can theoretically be achieved while still maintaining the reality of the horizon $r_+$. We can also find a set of parameters that satisfies $r_c > r_+$. 
\begin{figure}
\center
\includegraphics[width=.5\textwidth]{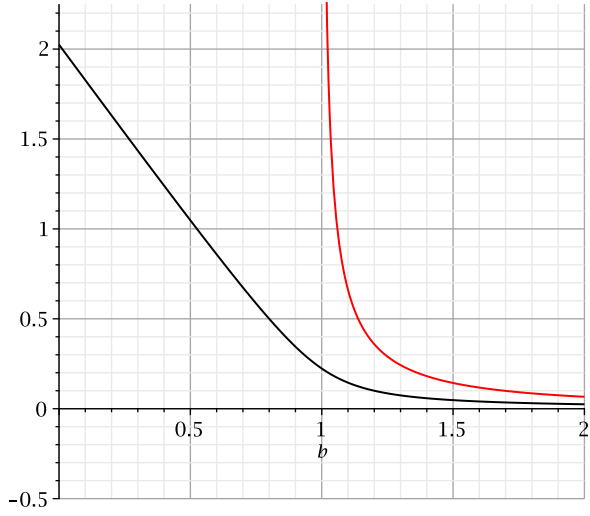}
\caption{Plot of $r_c$ (red) and the outter horizon $r_+$ (black) with varying values of $b$ with $M=1$, $l=0.3$, and $a=0.2$.}
\label{fig:rc-plot}
\end{figure}
As an example, figure \ref{fig:rc-plot} shows the plot of $r_c$ alongside the outter horizon $r_+$ with varying $b$, where the spacetime parameters being used are $M=1$, $l=0.3$, and $a=0.2$.
This means there could be a ring singularity at the equator with $r=r_c$ outside the horizon $r_+$ if $M<b$. This could be another condition for the "cosmic censorship hypothesis" to be aware of in KSTN spacetime other than the $r_+ \in \mathbb{R}$ condition, at least for $x=0$ case. Also, it can be seen that, in a case of extremal KSTN black hole, the horizon radius $r_+$ is negative if $M<b$ is satisfied. Although as interesting as it is, further discussion on the nature of the singularities in this spacetime is outside the scope of this paper.


\section{Conclusion}
In this paper, we obtain another conserved quantity known as the Carter constant by separating the Hamilton-Jacobi Equation for the Kerr-Sen-Taub-NUT spacetime. This metric was obtained and its properties were originally studied in \cite{Siahaan_2020_KSTN}.

In Sect. \ref{sec:ECPO}, we also searched for equatorial circular photon orbits (ECPO) utilizing the Carter constant obtained in Sect. \ref{sec:HJE-EOM}. In the process, we instead found another ring singularity that could appear outside the horizon $r_+$ if $M<b$, which is needed for the ECPO to be outside the horizon as well.
Analyzing the geodesics in KSTN spacetime as well as examining the Banados-Silk-West \cite{Ba_ados_2009,Zakria_2015} effect in KSTN spacetime are the future projects that can be done related to the KSTN solution.



\newpage
\appendix
\section{Additive separation of variables algorithm}\label{app}

Here I will discuss the algorithm used to do an additive separation of variables on a multivariate function, in particular, a function that depends on two variables. Suppose we have the function $F(x,y)$ which we want to rewrite as
\begin{equation}
F(x,y) = f(x) + g(y).
\end{equation}
As the author of \cite{Cherniavsky_2011} showed, the relation
\begin{equation}
\partial_x \partial_y F(x,y) = \partial_y \partial_x F(x,y) = 0
\end{equation}
must be true. Using the fact that a function can be expressed as
\begin{equation}
f(x) = \int{ \frac{ \dd f(x)}{ \dd x} \dd x} + C,
\end{equation}
and
\begin{equation}
\begin{split}
\partial_x F(x,y) = \frac{ \dd f(x)}{ \dd x},\\
\partial_y F(x,y) = \frac{ \dd g(y)}{ \dd y},
\end{split}
\end{equation}
the function $F(x,y)$ can be written as
\begin{equation}
\label{eq-app}
F(x,y) = \int{ \partial_x F(x,y) \dd x } + \int{ \partial_y F(x,y) \dd y} + C
\end{equation}
where the first term and the second term are respectively functions that only depend on $x$ and $y$, achieving the additive separation of variables. The appropriate constant $C$ is chosen to make the relation (\ref{eq-app}) true.

\bibliography{refs}
\bibliographystyle{h-physrev}

\providecommand{\href}[2]{#2}

\end{document}